# Time-gated optical-parametric-amplification and field-resolved detection of mid-infrared free-induction decay signals

Sergio Revuelta[1,*], Gerard Faustino[1], Ka Fai Mak[1,2,3,†], Ferenc Krausz[1,2,4,5] and Nicholas Karpowicz[1]

[1]Max-Planck-Institute of Quantum Optics, Laboratory for Attosecond Physics, Garching bei Munchen, Germany

[2] Faculty of Physics, Ludwig Maximilians Universitat Munchen, Garching bei Munchen, Germany

[3] School of Optical and Electronic Information, Huazhong University of Science and Technology, Wuhan, China

[4] Institute for Precision Prevention, The University of Hong Kong, Hong Kong SAR, China

[5] Center for Molecular Fingerprinting, Frontiers Foundation, Budapest, Hungary

[*]sergio.revuelta@mpq.mpg.de; [†]kafai.mak@hust.edu.cn

**Abstract**

Infrared spectroscopy is a key tool to study complex structural dynamics in molecular systems. However, resolving weak or broad vibrational modes is challenging due to their rapid decay. Here, we demonstrate a technique to significantly extend the detectable lifetime - and thereby spectral resolution - of molecular signatures by employing synchronized time-gated optical parametric amplification of mid-infrared free-induction decay signals. Utilizing a Cr:ZnS laser system and a nonlinear $ZnGeP_2$ crystal, we achieve a 10-fold increase in the electric field strength of the infrared molecular signal, yielding a two-orders of magnitude enhancement in dynamic range and a 2-fold increase in the signal-to-noise ratio. Using isopropyl alcohol as a model system, we demonstrate that OPA-enhanced detection extends the observable window of vibrational modes up to 72 %. This approach allows for the reconstruction of high-fidelity molecular fingerprints that resolve complex vibrational relaxation pathways and weak structural dynamics, previously hidden within the background and noise, providing a pathway for new high-sensitivity and time-resolved studies of complex chemical structures.

## I. Introduction

Infrared (IR) spectroscopy provides us with vital information about interband electronic transitions in semiconductors [1] or vibrational transitions in molecules from the eigenmodes that exchange energy with an IR optical excitation [2,3]. Until recently, IR spectroscopy was mainly performed via Fourier Transform Infrared (FTIR) spectrometers, using thermal radiation sources and benefiting from their compactness, low cost and flexibility for a diversity of applications [4–6]. Over the last decade, the use of coherent IR laser sources operating at MHz repetition rates, combined with highly-sensitive electro-optical sampling (EOS) detection has provided access to broadband absorption spectra of molecular systems with unprecedented sensitivity [7,8]. This novel technology can directly resolve complex dynamics in the time-domain, by capturing the free-induction decay (FID) in the wake of an ultrabrief excitation. Sampling the electric field oscillations of the exponentially-decaying FID signal delivers information about the vibrational eigenfrequencies of the molecules in the sample [9].

By capturing the temporal evolution of these IR-FID signals, we can obtain fingerprint-like information of the sample's molecular composition [10]. Changes of these signals constitute a sensitive probe to minuscule changes in molecular composition, rendering the technique useful for fundamental physical-chemistry [11,12] or applied clinical diagnostics[10,13] likewise. The approach holds promise for accessing tissue leakage proteins in cell-free blood samples, which are out of reach for conventional IR spectroscopy due to its limited dynamic range and the overwhelming background of highly abundant proteins [14,15].

In this work, we enhance the IR-FID EOS signals by employing synchronized time-gated optical-parametric-amplification (OPA) in order to amplify the weak, fast-decaying FID signals later in time, boosting the molecular ringing directly into the dynamic range of our electro-optical sampling detector. The IR-FID signals are amplified in the time-domain, concentrating the energy transferred from the pump pulse to the field cycles being detected, increasing the signal-to-noise ratio (SNR) and dynamic range (DR) by orders of magnitude. From the analysis of isopropyl alcohol as sample material, the unamplified IR-FID signal can be detected within a time window extending slightly more than 4 ps after IR excitation peak, with subsequent signal falling below the noise floor. Our time-gated amplification scheme increased the observable window of the C-C and C-O stretching modes by 72 and 66 % respectively, increasing the molecular information content of the signal owing to a larger number of observable field oscillations (and corresponding increase in spectral resolution). The demonstration of time-extended field-resolved detection will have positive implications for studies of ultrafast molecular dynamics by recovering weak, long-lived vibrational coherences previously buried beneath the instrument noise floor.

## II. Experimental concept

### Experimental setup

A schematic diagram of the measurement system is shown in Figure 1. A Cr:ZnS laser system providing a broad spectrum spanning from 1.4-2.8 um, sub-10 fs pulse width at 25 MHz repetition rate [16] was used to generate, amplify, and detect the MIR fields.

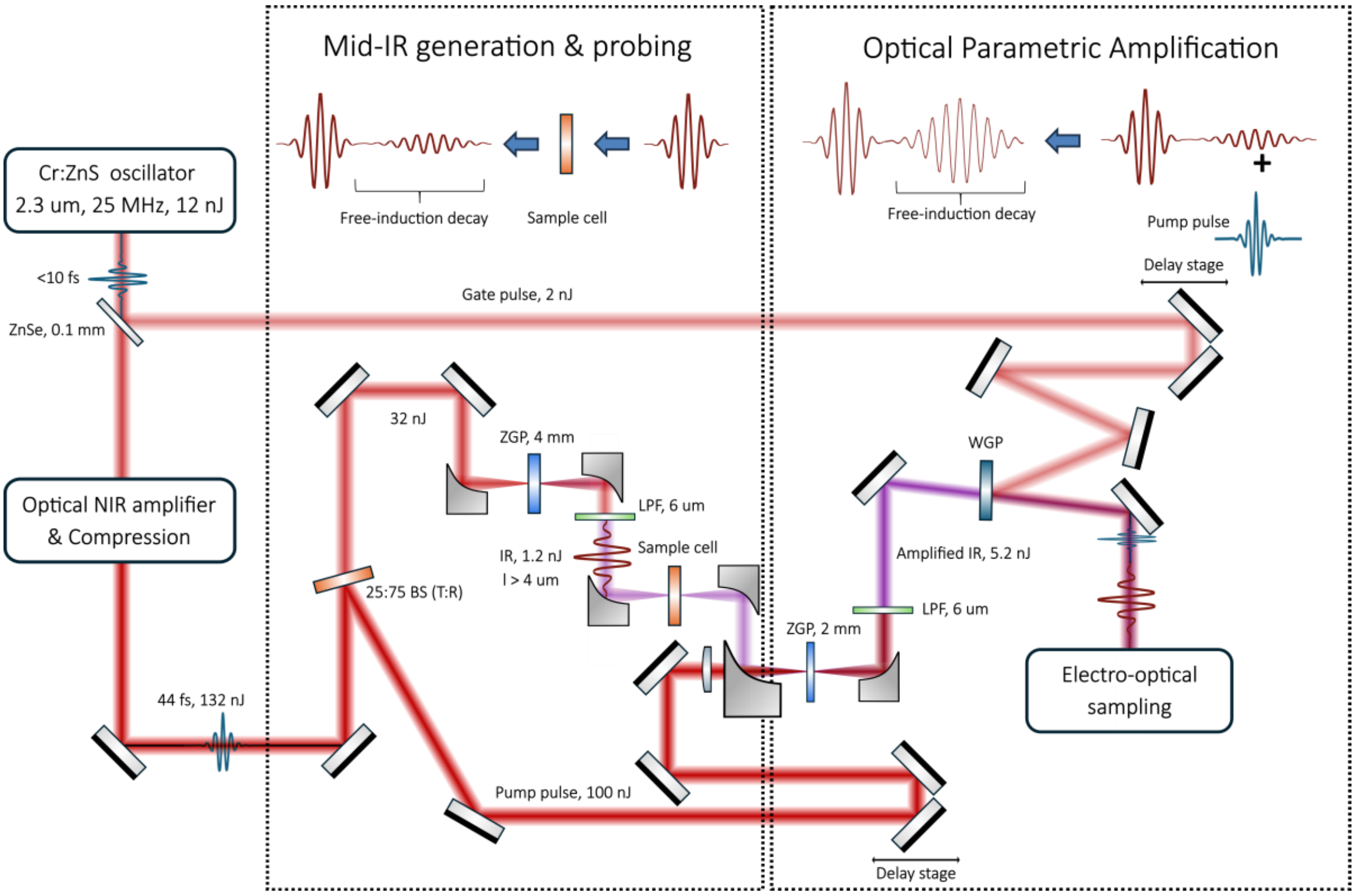

**Fig. 1.** A Cr:ZnS laser oscillator provides 12 nJ of <10 fs pulses, Δλ=1.4-2.8 µm and 25 MHz of repetition rate. The laser beam is split in two parts with a thin ZnSe (0.1-mm thickness) plate. The reflection (< 2 nJ) is used as a gate for electro-optical sampling. The transmitted pulse seeds a home-built single pass NIR amplifier. The 132 nJ of the amplified 2 µm light is split with a beam splitter with a 25:75 transmission:reflection ratio. The transmitted pulse is used to generate mid-IR light in a 4-mm-thick $ZnGeP_2$ crystal. OPA takes place in a 2-mm-thick $ZnGeP_2$ crystal, pumped by the 100 nJ reflected by the beam splitter. The MIR light is combined with the gate pulse for EOS using a wire-grid polarizer.

The laser light is split at the output of the laser with a 0.1-mm-thick ZnSe plate. The reflected part is used to sample the MIR light via EOS. The transmitted part is guided into a multistage NIR amplifier pumped with a 1.9 µm Thulium-doped fiber laser (TLR-120-WC), amplifying the input beam by a factor of ~30, from 155 mW to 3.3 W of average power (see Figure S1 in the Supplemental Material for details). To compress the 2 µm pulses after its amplification, we used a combination of bulk dispersion (20 mm Sapphire + 5 mm CaF2) and multilayer chirped mirrors. The pulses were compressed from 230 fs to 44 fs (for details see Figures SF2 in the Supplemental Material).

The compressed 2 µm beam is guided into a beam 25:75 splitter. The transmitted 25% is focused into a 4-mm-thick $ZnGeP_2$ (ZGP, DIEN TECH Co., Ltd) crystal for frequency-down conversion via intra-pulse difference frequency generation (IPDFG), to generate MIR radiation [17]. This crystal was chosen for the MIR generation and OPA for its high non-linear coefficient, broadband transparency and high-energy tunability in the MIR regime [18–21].

The field reflected from the beam splitter is used as the pump beam for the OPA. The pump passes through a delay line to control the temporal overlap between the MIR and the 2 µm pump. The MIR is focused into a ZGP crystal 2-mm-thick (DIEN TECH Co., Ltd) with an off-axis parabolic mirror. The pump is focused with an AR-coated 150 mm lens and send directly to the crystal through a hole in the parabolic mirror (see Figure 1).

To retrieve the time-domain traces of the MIR electric field, we recorded the waveform via EOS, using the reflection of the thin-film ZnSe as the gate (see Figure 1) as the gate pulse. The latter and the MIR light are combined with a $BaF_2$ wire-grid polarizer. Both beams are focused into a 0.1-mm-thick GaSe crystal z-cut for type-II phase matching [17,22].

## III. Results and discussion

### Optical parametric amplification of mid-IR light

We characterize the optical amplification of the MIR light in ambient conditions (i.e., without a sample and without purging of the setup). Figure 2a shows the result obtained from the EOS measured without the OPA (solid blue line). The translation stages for EOS and OPA are moved synchronously to amplify the measured section of the electric field. The resulting signal can be seen in Figure 2a (solid blue line). Here we obtain an amplification of 10 for the MIR electric field strength. To quantify the limits of detection of our measurement setup, we evaluate, for the unamplified MIR, the signal-to-noise ratio (SNR) and dynamic range (DR) at different time positions of 0 and 2.71 ps (vertical lines on Figure 2b). At 0 ps, we obtain a SNR of 62 and a DR of 95 dB, while at 2.71 ps the SNR and DR are 25 and 65 dB respectively (see Figure S3 and reference [23] for details).

With the OPA, the SNR and DR was 133 and 115 dB at 0 ps and 51 and 85 dB at 2.71 ps [23], corresponding to a 2-fold increase of the SNR and a 2-orders of magnitude increase of the DR with the optical amplification via OPA of the MIR light. In terms of average power, when the pump overlapped with the seed pulse, the enhancement of the average power was a factor of 4.6, from 28 to 129 mW.

Figure 2b shows the Fourier-transform of the full EOS trace without amplification. The resulting spectrum matches with that measured independently by a calibrated Fourier-transform infrared spectrometer (FTIR, ARCoptix FTMIR-L1-160-DLA). Figure 2c shows the spectra corresponding to the EOS traces with and without OPA amplification (from Figure 2a).

As can be seen the measured mid-infrared spectral intensity is increased by more than one order of magnitude with the OPA, revealing spectral content that was previously below the noise floor. This magnitude of increase is consistent with our numerical simulations (see Figure S4 in the Supplemental Material and reference [24]).

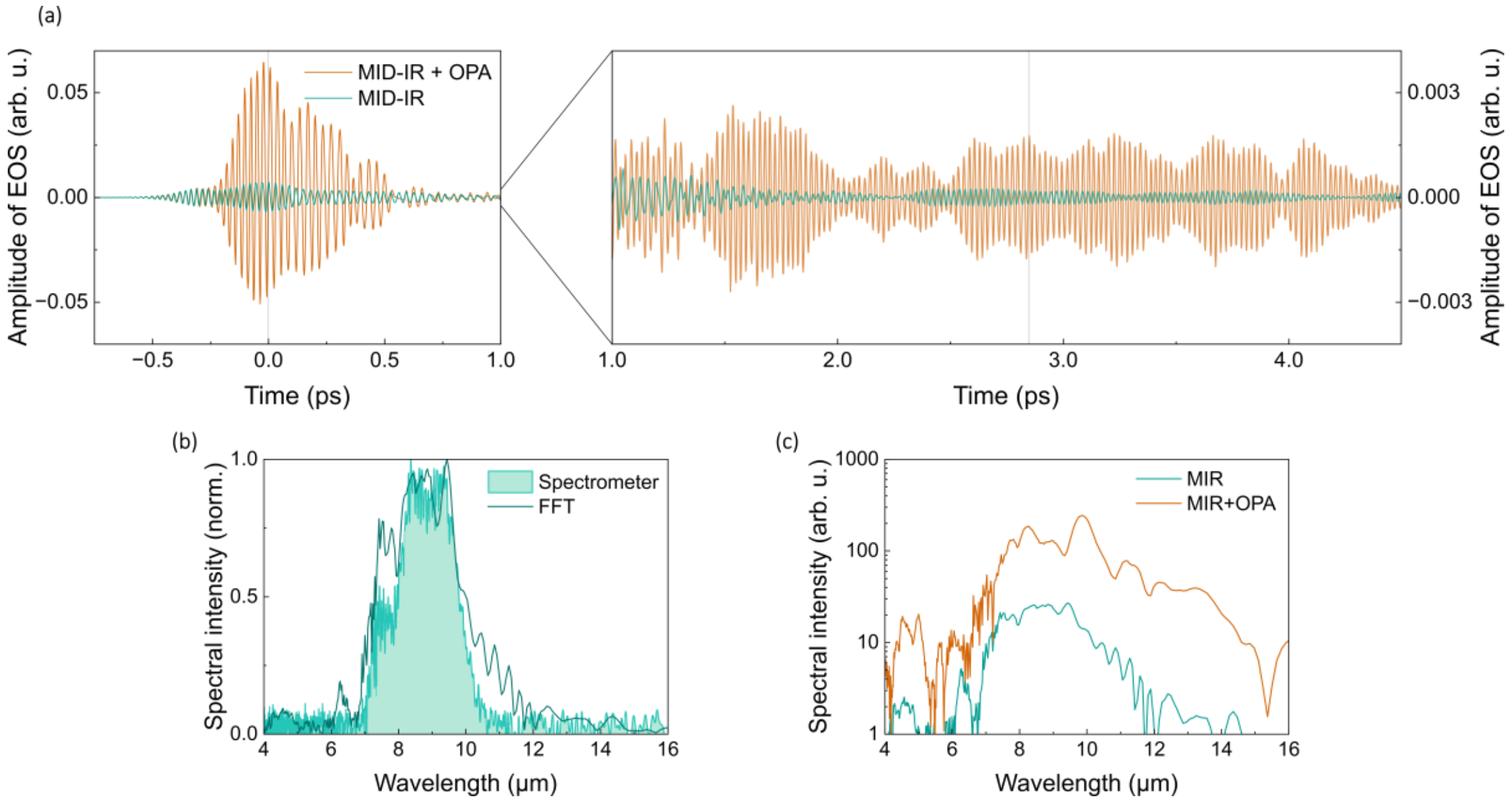

**Figure 2.** (a) Time-domain electric field traces of the mid-IR light without (solid blue line) and with (red solid line) optical-parametric-amplification. Vertical lines refer to the time position used for SNR and DR acquisition. (b) Fourier transform of the mid-IR EOS trace without optical parametric amplification together with the retrieved spectra from the FTIR spectrometer. (c) Fourier transform of the mid-IR EOS traces without (solid blue line) and with (solid red line) optical parametric amplification.

## Optical parametric amplification of field-induction decay signals in isopropyl alcohol

The above-mentioned results show the capabilities of improved EOS detection of mid-IR light via synchronized OPA. As a proof-of-principle of its capabilities in spectroscopy, we employed the setup to probe and amplify the FID signals of liquid isopropyl alcohol after MIR excitation (2-Propanol >99.8% min, Carl Roth 6752.4). Specifically, we were interested in resolving the FID signals from the strong modes of C-O stretching (1075-1150 cm$^{-1}$ / 8.7 - 9.3 µm / 32.2 – 34.5 THz) and C-C stretching (950 cm$^{-1}$ / 10.5 µm / 28.5 THz) [25,26] that fall within our spectral window.

The liquid isopropyl alcohol was placed in a sample cell with 3-mm thick ZnSe windows on either side and the whole setup was purged with nitrogen gas. This is to avoid atmospheric water droplets, in particular those between the OPA stage and the EOS stage, from causing measurement artifacts. Figure 3a shows the EOS trace of the MIR light passing through the sample without (solid blue line) and with (solid red line) OPA. As before, we observe an increase of the electric field strength by a factor of 10 and an amplification of the spectral intensity by more than 1 order of magnitude (see Figure SF5 in the Supplemental Material for the FFT of the EOS traces and the spectrums recovered by the FTIR).

To evaluate the temporal evolution of the signal's spectral content, spectrograms of the EOS traces (Figure 3a) were retrieved using the Gabor transform method. Figure 3b shows the spectrogram of the EOS signal for the unamplified beam, normalized to the background noise (see Supplementary Material, Figure SF7). Both the unamplified (Figure 3b) and OPA-amplified (Figure 3c) spectrograms are dominated by an intense feature centered at 30 THz, corresponding to the MIR excitation pulse, which is slightly chirped. Following this initial pulse, distinct spectral lines emerge at 28.5 and 34.5 THz (indicated by horizontal dashed lines), which are assigned to the C–C and C–O stretching modes, respectively.

To quantify the impact of the OPA, Figures 3d and 3e compare the temporal traces of these modes without (solid blue lines) and with amplification (solid red lines). Notably, OPA amplification extends the observable window of the C–C stretching mode from 4.78 ps to 8.24 ps, a 72 % increase. Similarly, the observable duration of the C–O mode increases by 66 %, extending from 4.39 ps to 7.29 ps.  Since the coherent signal of many vibrational modes last only a few hundred of femtoseconds [27,28], the ability to lift these above the detection limit opens the door to study structural dynamics and hidden weak-mode-coupling channels. This is a critical capability for avoiding artificial spectral broadening and resolving narrow vibrational modes in complex media.

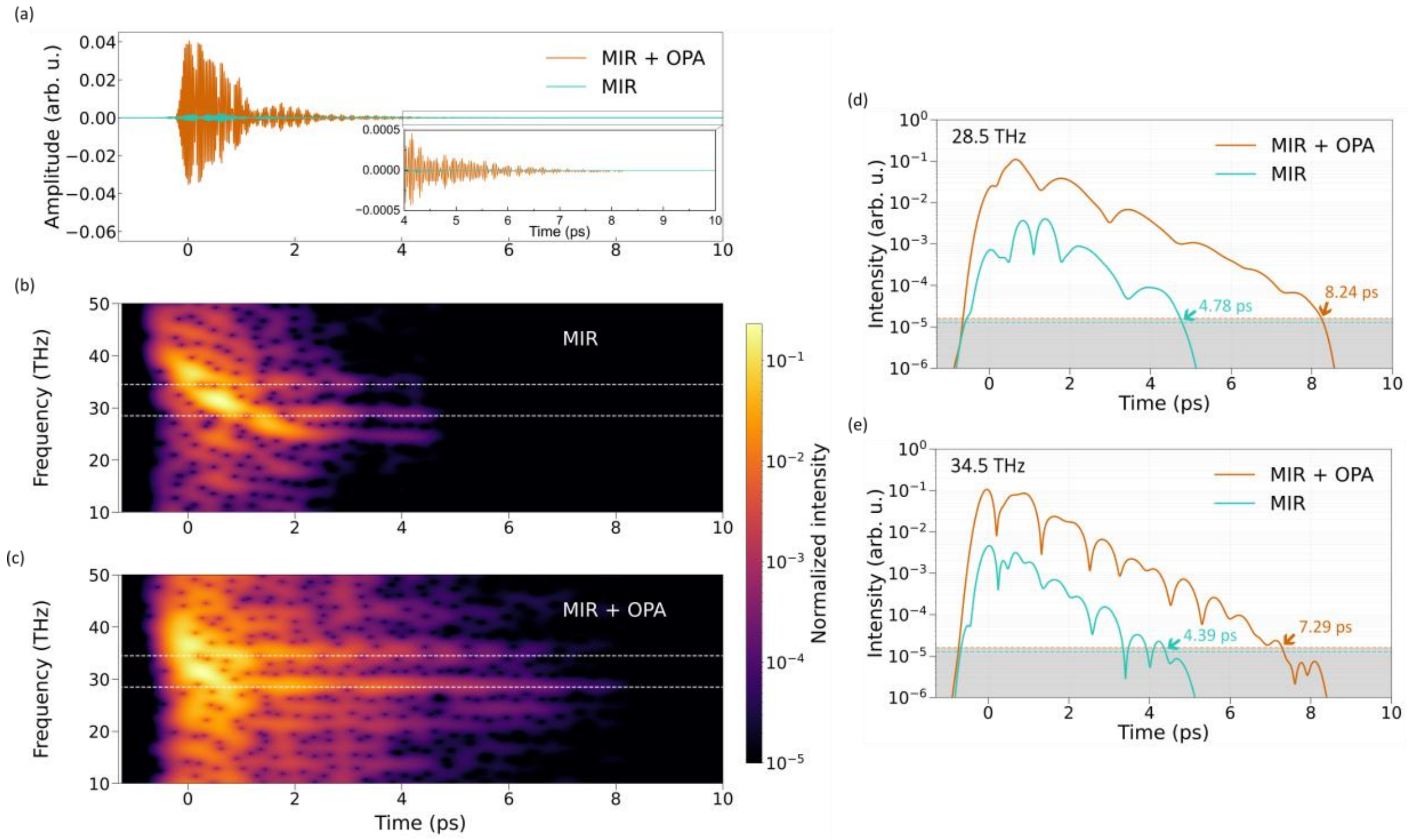

**Figure 3.** (a) Time-domain electric field traces of the mid-IR light without (solid blue line) and with (red solid line) optical-parametric-amplification after transmission through 2-Propanol. Inset: Zoom-in view of the oscillations at the edge of the pulse. (b) and (c) Spectrograms obtained by Gabor transformation of the EOS trace in (a) without and with OPA respectively. Horizontal dashed lines at 28.5 and 34.5 THz are shown to indicate the positions of the C-C and C-O modes. (d) and (e) frequency-slice at 28.5 and 34.5 THz as indicated, for the MIR and MIR+OPA. Solid dashed lines indicate the background noise for each spectrogram (see Supplemental Material for the noise calculation).

## IV. Conclusions and outlook

In conclusion, we have demonstrated a high-repetition-rate, time-gated spectroscopic instrument that pushes the frontiers of MIR spectroscopy in terms of sensitivity and information content via optical parametric amplification. By amplifying infrared free-induction decay responses in the time domain, we achieved a dynamic range enhancement of over two orders of magnitude alongside a 2-fold increase in the signal-to-noise ratio. Crucially, this approach extends the observable coherence time-window for vibrational resonances, specifically yielding an extension of 72% and 66% for the C–C and C–O stretching modes respectively in isopropyl alcohol. This extension is critical for isolating rapidly-decaying free-induction decay signals from the background generated by the excitation pulse.

The capability of OPA to convert high-intensity pump photons into signal gain offers a unique way to enhance the sensitivity and specificity of molecular fingerprinting. Scaling the pump power along with advanced engineering of ZGP crystals (for avoiding losses via phonon absorption and impurities) [29] holds promise for a powerful tool for high-dynamic-range sensing in complex chemical environments where relevant molecular response comes in combination with a strong background of dominant solvent or matrix signals. In the realm of field-resolved nonlinear spectroscopy, where the intensity of excitation on the sample may determine the observed dynamics, post-amplification in the detection process provides an opportunity to normalize the detected signal for a range of intensities in the sample. We believe these results provide a clear framework for extending the observable time window and dynamic range of field-resolved measurements.

**Acknowledgements**

We acknowledge financial support from the cooperation program of the Fraunhofer Institute for industrial mathematics ITWM and the Max Planck Institute of Quantum Optics (project iFRS). We are grateful to Alexander Weigel (Ludwig-Maximilians-Universität München), Volodymyr Pervak (Ludwig-Maximilians-Universität München) and Ioachim Pupeza (RPTU University Kaiserslautern-Landau) for fruitful discussions on mid-IR generation, dispersive mirrors and IR-spectroscopy.

## References

[1] J. Toudert and R. Serna, Interband transitions in semi-metals, semiconductors, and topological insulators: a new driving force for plasmonics and nanophotonics, Opt. Mater. Express **7**, 2299 (2017).

[2] M. A. Czarnecki, Y. Morisawa, Y. Futami, and Y. Ozaki, Advances in Molecular Structure and Interaction Studies Using Near-Infrared Spectroscopy, Chem. Rev. **115**, 9707 (2015).

[3] L. M. Ng and R. Simmons, Infrared Spectroscopy, Anal. Chem. **71**, (1999).

[4] P. Griffiths and J. A. De Haseth, *Fourier Transform Infrared Spectrometry* (Wiley & Sons Ltd, 2007).

[5] P. I. , Haris and A. Barth, *Biological and Biomedical Infrared Spectroscopy |* (IOS Press, 2009).

[6] C. Kötting and K. Gerwert, Proteins in action monitored by time-resolved FTIR spectroscopy, ChemPhysChem **6**, 881 (2005).

[7] M. T. Peschel, M. Högner, T. Buberl, D. Keefer, R. de Vivie-Riedle, and I. Pupeza, Sub-optical-cycle light-matter energy transfer in molecular vibrational spectroscopy, Nature Communications 2022 13:1 **13**, 5897 (2022).

[8] C. Hofer, D. Gerz, M. Gebhardt, T. Heuermann, T. P. Butler, C. Gaida, J. Limpert, F. Krausz, and I. Pupeza, *Electro-Optic Sampling with Percent-Level Detection Efficiency*, in *2021 Conference on Lasers and Electro-Optics Europe and European Quantum Electronics Conference (2021), Paper Cf_6_5* (Optica Publishing Group, 2021), p. cf_6_5.

[9] I. Coddington, W. C. Swann, and N. R. Newbury, Time-domain spectroscopy of molecular free-induction decay in the infrared, Opt. Lett. **35**, 1395 (2010).

[10] I. Pupeza et al., Field-resolved infrared spectroscopy of biological systems, Nature **577**, 52 (2020).

[11] N. T. Hunt, 2D-IR spectroscopy: Ultrafast insights into biomolecule structure and function, Chem. Soc. Rev. **38**, 1837 (2009).

[12] A. A. Bunaciu, Ş. Fleschin, V. D. Hoang, and H. Y. Aboul-Enein, Vibrational Spectroscopy in Body Fluids Analysis, Crit. Rev. Anal. Chem. **47**, 67 (2017).

[13] K. V. Kepesidis et al., Electric-Field Molecular Fingerprinting to Probe Cancer, ACS Cent. Sci. **11**, 560 (2025).

[14] P. E. Geyer, L. M. Holdt, D. Teupser, and M. Mann, Revisiting biomarker discovery by plasma proteomics, Mol. Syst. Biol. **13**, (2017).

[15] M. Liu, R. M. Gray, R. Arkadev, L. Ledezma, and A. Marandi, Optical-parametric-amplification-enhanced background-free spectroscopy, Optics Letters, Vol. 49, Issue 11, Pp. 2914-2917 **49**, 2914 (2024).

[16] M. Kowalczyk, N. Nagl, P. Steinleitner, N. Karpowicz, V. Pervak, A. Głuszek, A. Hudzikowski, F. Krausz, K. F. Mak, and A. Weigel, Ultra-CEP-stable single-cycle pulses at 2.2 µm, Optica **10**, 801 (2023).

[17] P. Steinleitner et al., Single-cycle infrared waveform control, Nat. Photonics **16**, 512 (2022).

[18] Z. Lv, Y. Shen, Y. Wen, E. P. Wang, Z. M. Wang, W. L. Li, Y. Bo, and Q. J. Peng, High power widely tunable mid-IR (5–7.2 µm) ZnGeP2 optical parametric oscillator pumped by a 2.09 µm laser, Infrared Phys. Technol. **134**, (2023).

[19] D. Wei, M. Wu, J. Tang, W. Cheng, X. Hua, J. H. Li, B. Yao, T. Dai, X. Duan, and Y. Ju, High-efficiency hundred-picosecond mid-infrared ZnGeP2 OPG/OPA system for mJ-level pulse generation, Opt. Laser Technol. **180**, (2025).

[20] D. Wei, J. Tang, X. Hua, J. Li, M. Wu, W. Cheng, B. Yao, J. Yao, T. Dai, and X. Duan, Narrowband mid-infrared optical parametric generator based on a type-II BaGa4Se7 crystal, Opt. Lett. **49**, 4278 (2024).

[21] A. Berrou, M. A. Medina, M. Piotrowski, V. Allheily, and A. Hildenbrand-Dhollande, High-powered non-planar ring resonator mid-IR ZGP OPO, a parametric study, Optics Continuum **4**, 1433 (2025).

[22] S. A. Hussain, C. Hofer, M. Högner, W. Schweinberger, T. Buberl, D. Bausch, M. Huber, F. Krausz, and I. Pupeza, Sub-attosecond-precision optical-waveform stability measurements using electro-optic sampling, Sci. Rep. **14**, (2024).

[23] M. Naftaly and R. Dudley, Methodologies for determining the dynamic ranges and signal-to-noise ratios of terahertz time-domain spectrometers, Opt. Lett. **34**, 1213 (2009).

[24] N. Karpowicz, Open source, heterogeneous, nonlinear optics simulation, Optics Continuum **2**, 2244 (2023).

[25] P. J. Linstrom and W. G. Mallard, The NIST Chemistry WebBook: A chemical data resource on the Internet. Database Number 69, J. Chem. Eng. Data **46**, 1059 (2001).

[26] M. Kennema, I. B. D. De Castro, F. Meemken, and R. Rinaldi, Liquid-Phase H-Transfer from 2-Propanol to Phenol on Raney Ni: Surface Processes and Inhibition, ACS Catal. **7**, 2437 (2017).

[27] A. M. Stingel and P. B. Petersen, Couplings Across the Vibrational Spectrum Caused by Strong Hydrogen Bonds: A Continuum 2D IR Study of the 7-Azaindole–Acetic Acid Heterodimer, J. Phys. Chem. B **120**, 10768 (2016).

[28] E. T. J. Nibbering and T. Elsaesser, Ultrafast Vibrational Dynamics of Hydrogen Bonds in the Condensed Phase, Chem. Rev. **104**, 1887 (2004).

[29] P. G. Schunemann, K. T. Zawilski, J. Jesenovec, and E. D. Readinger, Mid-infrared nonlinear optical materials for ultrafast sources: opinion, Opt. Mater. Express **16**, 656 (2026).

Supplemental Material

# Optical NIR amplifier and compression

The custom-made multistage amplifier and compression consists on 4 different amplification stages and 1 part related with GDD and TOD compensation (see Figure S1). Each one amplification stage is based on a Cr:ZnS crystal as gain media. The crystals are pumped with a 1.9 µm Thulium-doped fiber laser (IPG Photonics, TLR-120-WC). To avoid thermal damage the crystals are cooled to 8°C.

After the amplification, the 2 µm beams needs to be compressed for an efficient MIR generation and OPA. Figure S1 shows a scheme of the compression part. To compensate the group delay dispersion (GDD) and third-order dispersion (TOD) we used a combination of substrates with dispersive mirrors. The FROG traces related with the input and output of the compression part after its amplification can be seen in Figure S2.

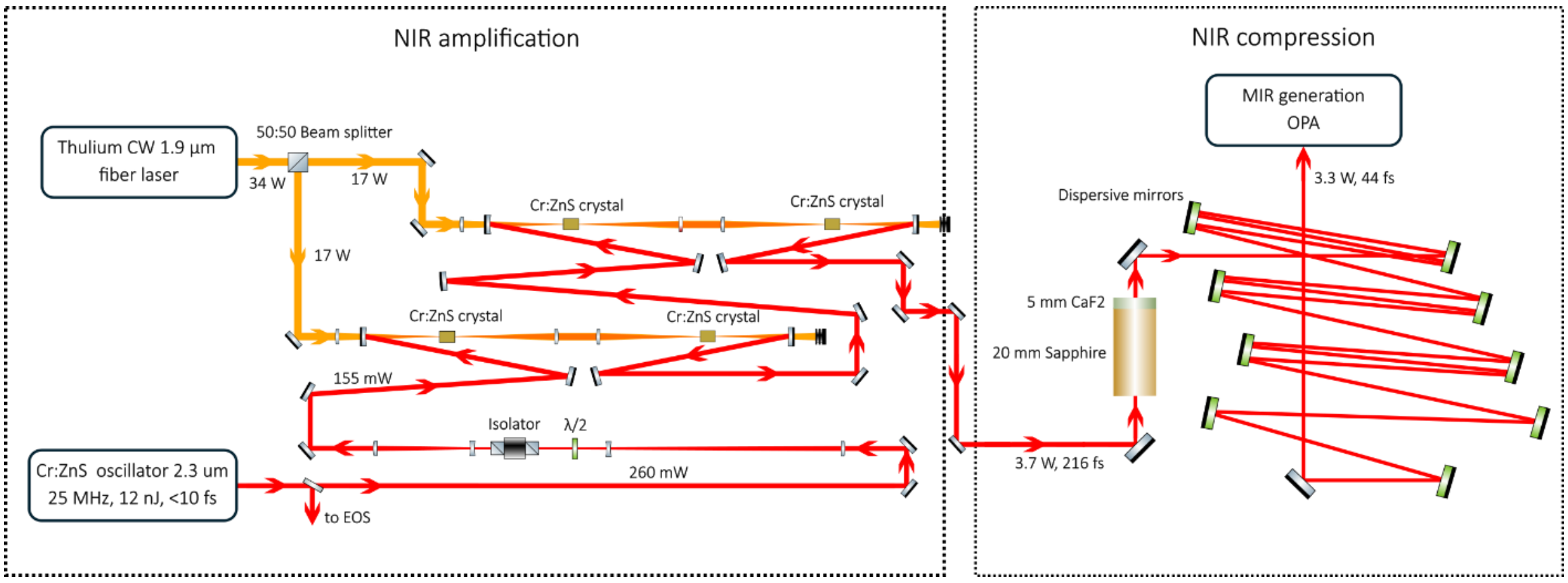

**Figure S1.** Scheme of the NIR amplifier and compression part. A train of ultra-short laser pulses centered 2.3 µm at 25 MHz and with a pulse duration of <10 fs enters the setup and splits in two different beams by a 0.1mm-thick ZnS film. The reflected part is being used for the EOS (see Figure SX). The beam diameter of the transmitted part is being reduced by a telescope to pass through a Faraday Isolator. The beam is being reflected by an curve input coupler to focus on the 1$^{st}$ Cr:ZnS crystal. A thulium 1.9 µm fiber laser operating in CW mode is used to pump the gain Cr:ZnS crystals. 50 % of the Thulium laser power is used to pump the 1$^{st}$ and 2$^{nd}$ crystals, while the rest is used to pump the 3$^{rd}$ and 4$^{th}$ crystals. After the amplification, the duration of the train laser pulses is 216 fs (see Figure S2). To compensate the GDD and TOD, we used a combination of glass materials (20 mm of Sapphire + 5 mm of CaF2) and custom-made mirrors to compress the pulses to almost its Fourier-Transform Limit (FTL=35 fs).

# FROG traces before and after compression

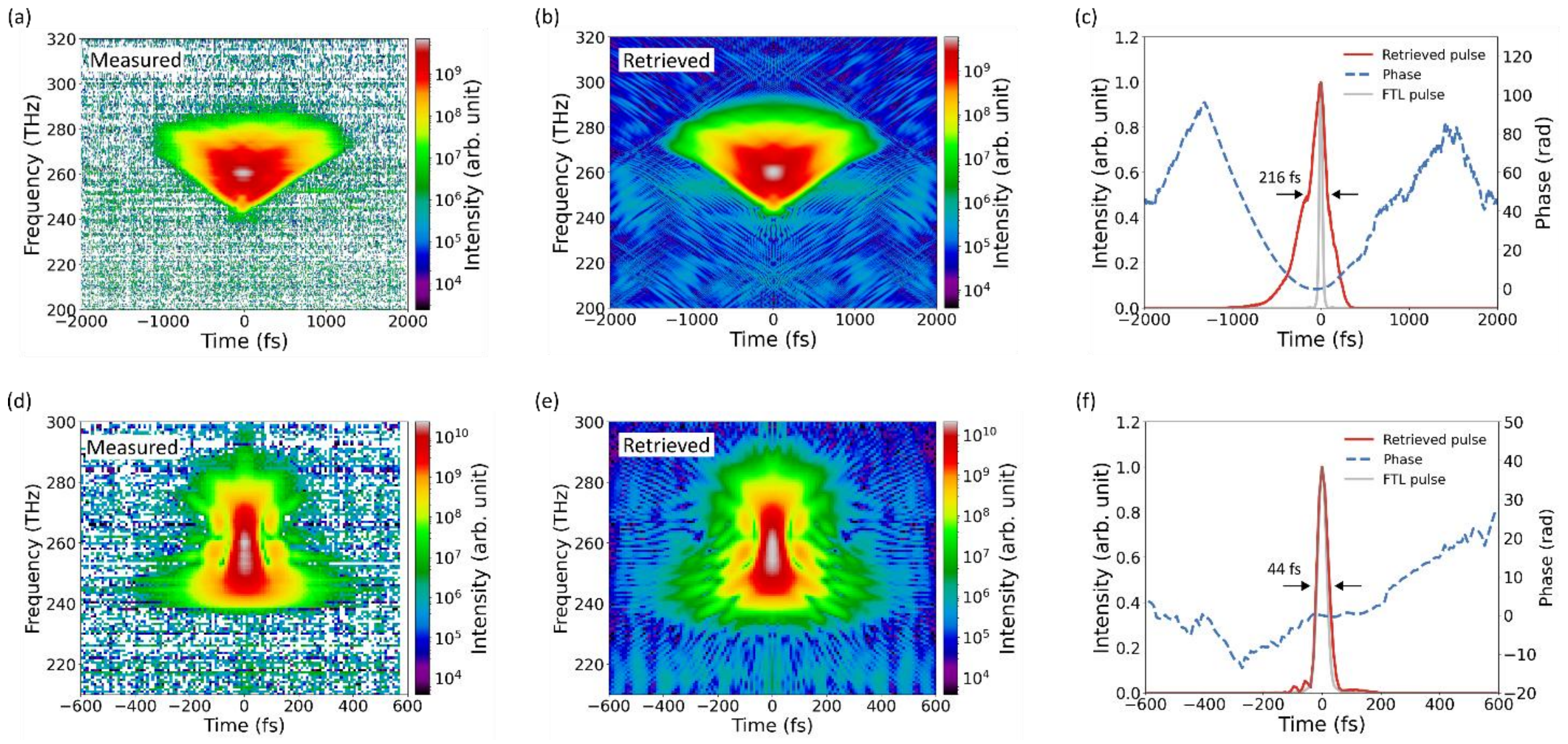

**Figure S2.** Top figures represent the pulse duration before the compression. (a) Measured FROG spectrogram. (b) Retrieved FROG spectrogram. (c) Temporal intensity profile (red curve) and temporal phase (blue dashed line) of the amplified pulse together with the Fourier-Transform limit (FTL = 35 fs) of the measured spectrum (grey line). Bottom figures represent the pulse duration after the compression. (d) Measured FROG spectrogram. (e) Retrieved FROG spectrogram. (f) Temporal intensity profiled (red curve) and temporal phase (blue dashed line) of the amplified and compressed pulse, together with Fourier-Transform limit (FTL = 35 fs) of the measured spectrogram (grey line).

# Signal-to-noise ratio (SNR) and dynamic range (DR) calculation

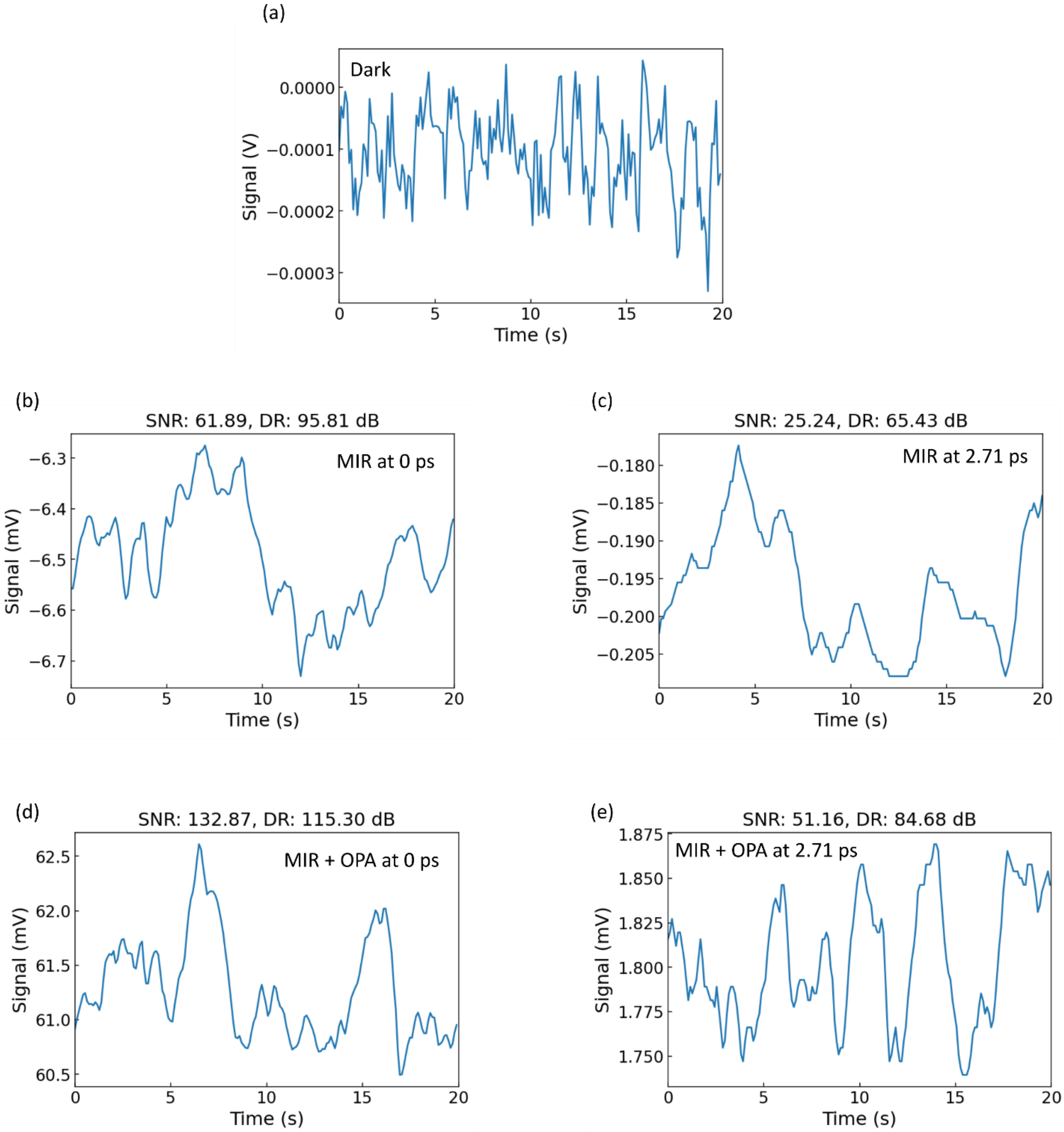

**Figure S3.** Measured signal during 20 seconds for the SNR and DR calculation. (a) Dark measurement (i.e., gate signal without MIR light). (b) MIR signal at 0 ps. (c) MIR signal at 2.71 ps. (d) MIR+OPA signal at 0 ps. (e) MIR+OPA signal at 2.71 ps. The SNR and DR was calculated following the methodology described in [1].

**Table S1.** Mean value and standard deviation extracted from Figure S3.

| | Mean value (V) | Standard deviation (V) |
|---|---|---|
| Dark | -1.052425E-07 | 6.727378E-08 |
| MIR (0 ps) | -6.495041E-03 | 1.049460E-04 |
| MIR (2.71 ps) | -1.966568E-04 | 7.791856E-06 |
| MIR+OPA (0 ps) | 6.127239E-02 | 4.611515E-04 |
| MIR + OPA (2.71 ps) | 1.803028E-03 | 3.524037E-05 |

# Numerical Simulations for OPA of MIR light

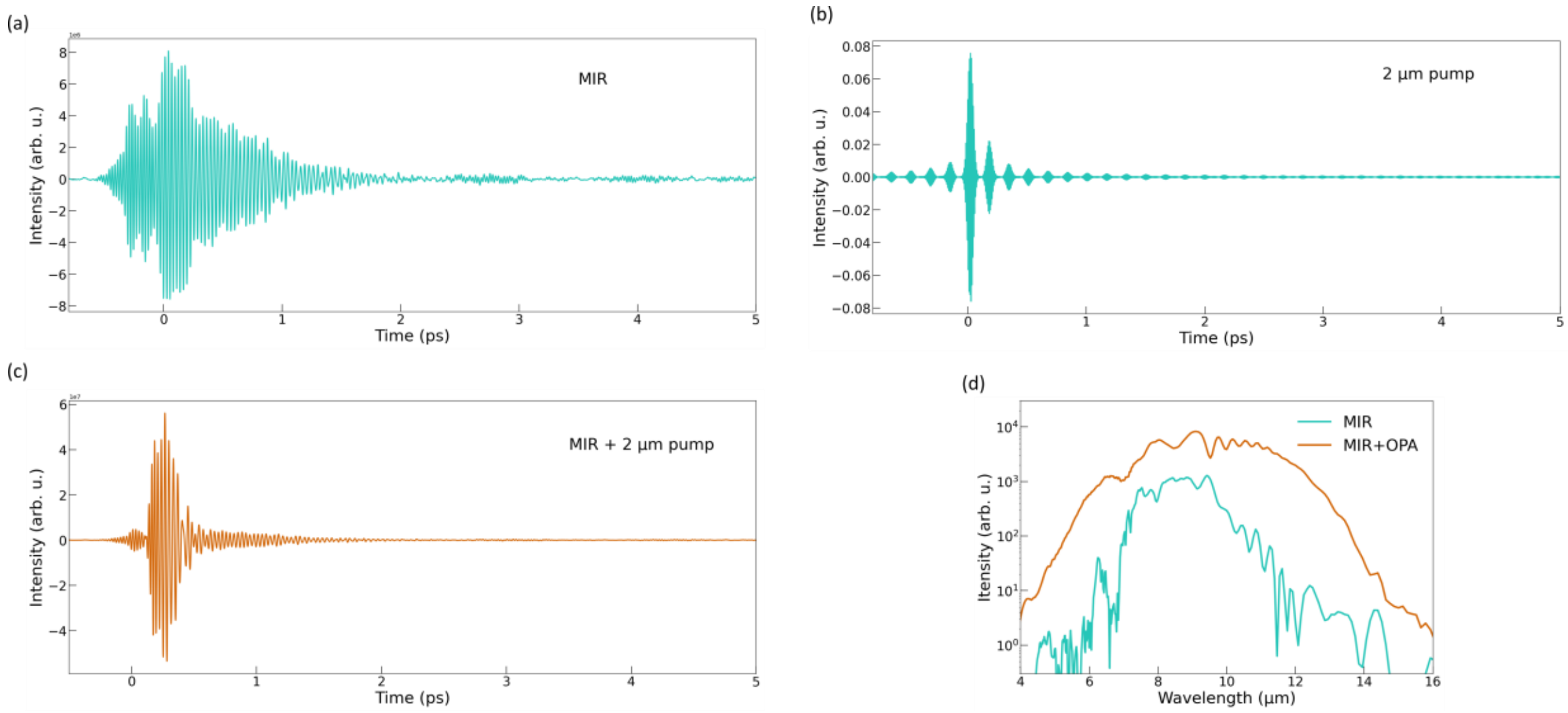

**Figure S4.** (a) MIR electric field without OPA from Figure 2a. (b) Electric field of the 2 µm pump beam, obtained from FROG results of Figure S2. (c) LightwaveExplorer [2] simulations of MIR amplification via OPA. Input parameters are the measured MIR electric field, reconstructed 2 µm pump beam and measured beam characterizations (see Figure S6). This waveform shows the highest amplification retrieved from simulations. (d) Fourier-transform of waveform shown in Figure S4c. 1 order of magnitude increase of the MIR light is achieved.

# Fourier-transform from EOS waveforms of Figure 3a

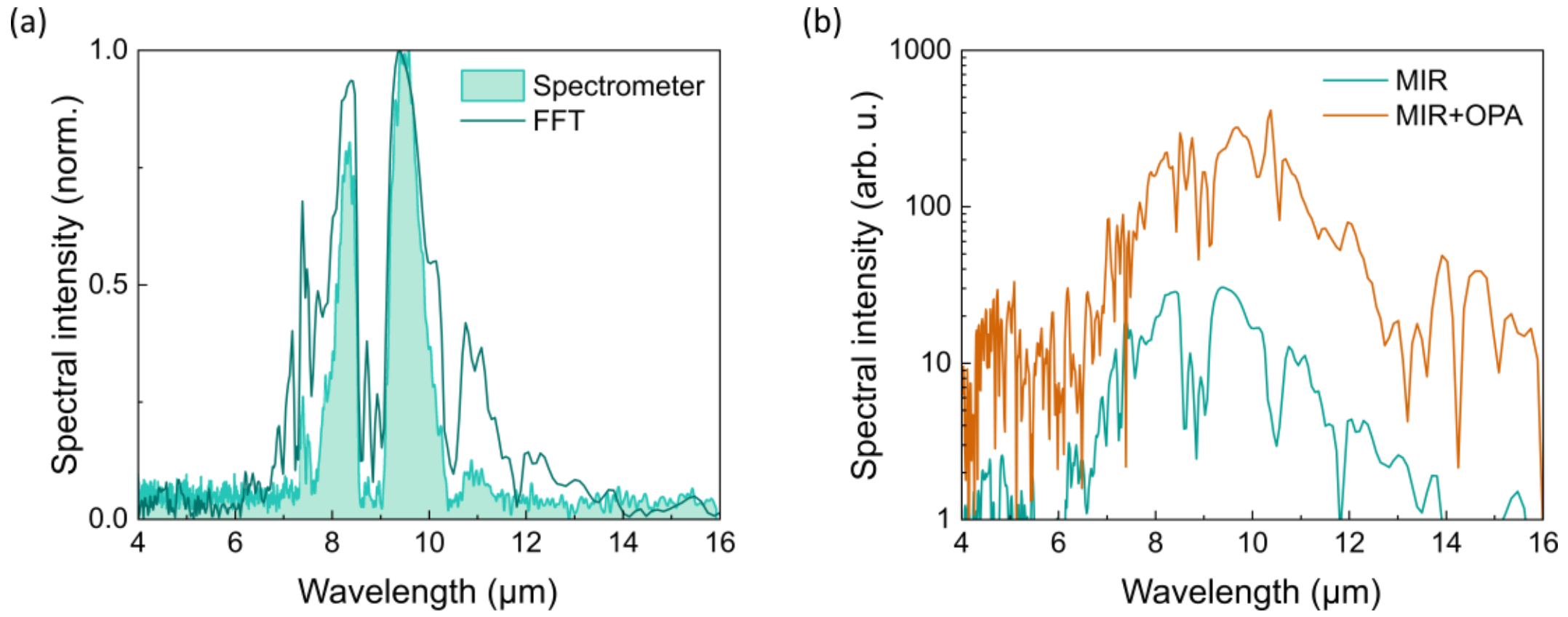

**Figure S5.** (a) Fourier transform of the mid-IR EOS trace shown in Figure 3a without OPA together with the retrieved spectra from the FTIR spectrometer. (b) Fourier transform of the MIR EOS traces shown in Figure 3a without (solid blue line) and with (solid red line) OPA.

# Experimental parameters for simulations of Figure S4

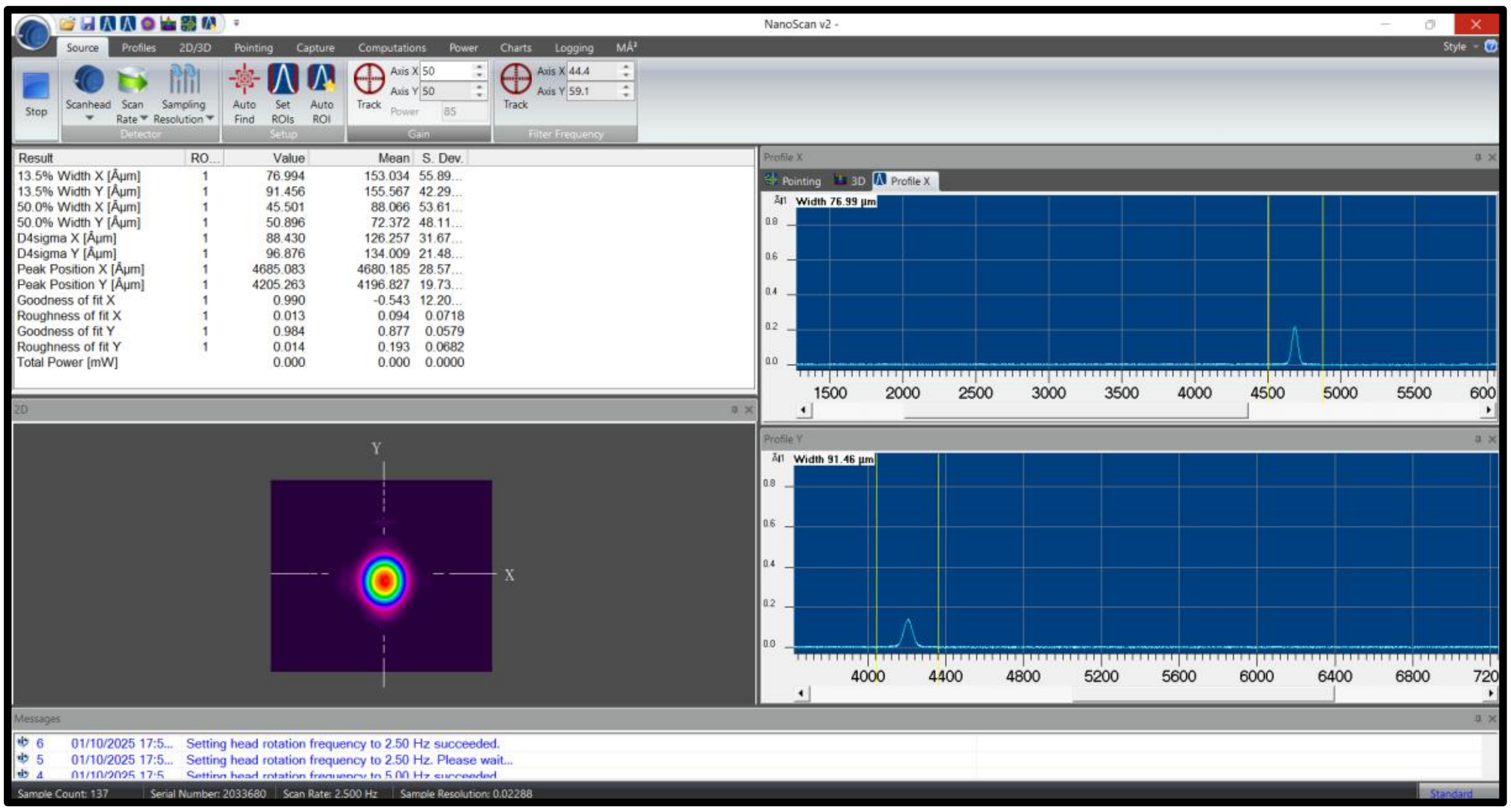

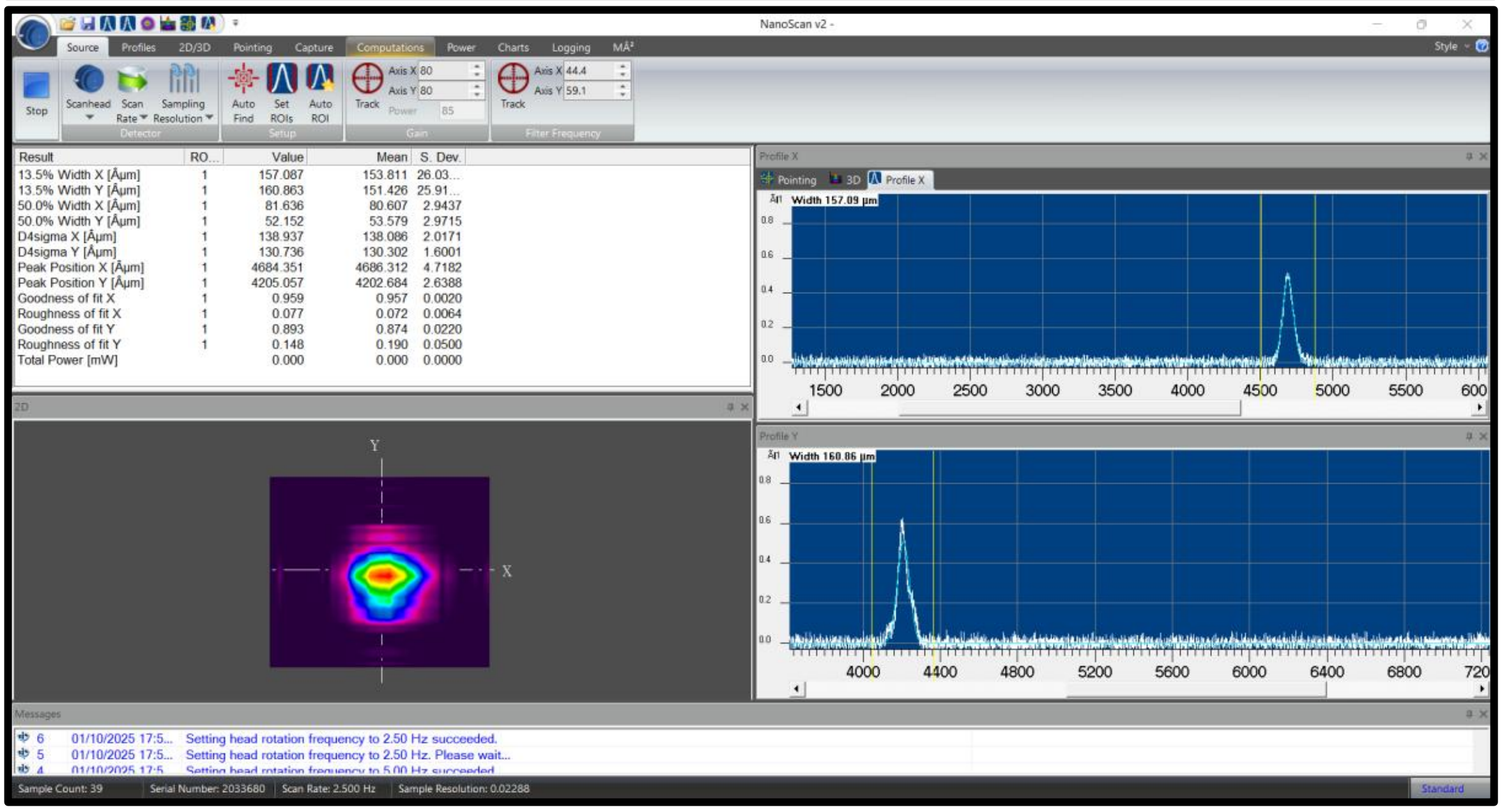

**Figure S6.** Top: Beam profile measured at the focus for the 2 µm pump beam at the OPA stage. The measured spot size is 88 and 97 µm for the X and Y axis respectively. Bottom: Beam profile measured at the focus for the MIR light at the OPA stage. The measured spot size is 139 and 130 for the X and Y axis respectively. Data was collected by the Nanoscan 2s Pyro/9/5.

# Spectrogram’s noise calculation

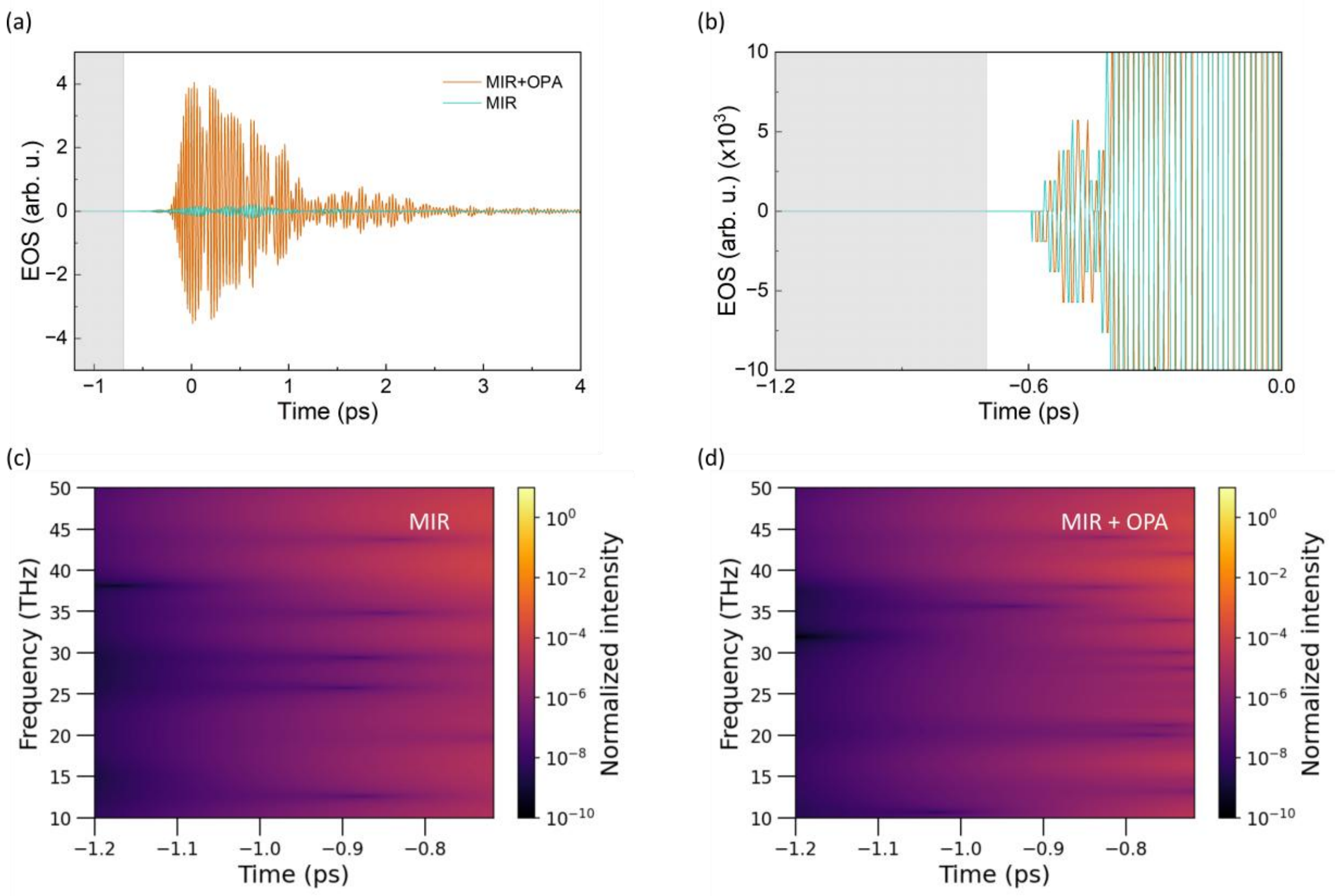

**Figure S7.** (a) and (b) time-domain signals presented in Fig 3a in the main text. Grey area refers to the isolated part where the noise was calculated for establishing a base line for the frequency slices shown in Fig.3d,e. (c) and (d) retrieved spectrograms for the time-domain traces of the MIR and MIR+OPA respectively. The obtained noises for the data with and without OPA are 1.59E-5 and 1.27E-5 respectively.

## References

[1] M. Naftaly and R. Dudley, Methodologies for determining the dynamic ranges and signal-to-noise ratios of terahertz time-domain spectrometers, Opt. Lett. **34**, 1213 (2009).

[2] N. Karpowicz, Open source, heterogeneous, nonlinear optics simulation, Optics Continuum **2**, 2244 (2023).